\documentstyle[12pt]{article}
\begin{document}

\begin{center}
{\large {\bf Altenative bi-Hamiltonian structures for \\ WDVV equations of
associativity }}\\[15mm] {\large J. Kalayc{\i}$^{*\dagger}$ and
Y. Nutku$^{*}$ }\\[2mm] $^*$ Feza G\"{u}rsey
Institute,  P. K. 6 \c{C}engelk\"{o}y 81220 Istanbul,
Turkey \\[2mm] $^\dagger$ Istanbul Technical University \\ Department of
Physics\\ 80626 Maslak, Istanbul, Turkey \\[3mm]
\today
\vspace{1 cm}
\end{center}

The WDVV equations of associativity in 2-d topological field theory are
completely integrable third order Monge-Amp\`ere equations which admit
bi-Hamiltonian structure. The time variable plays a distinguished role in
the discussion of Hamiltonian structure whereas in the theory of WDVV
equations none of the independent variables merits such a distinction. WDVV
equations admit very different alternative Hamiltonian structures under
different possible choices of the time variable but all these various
Hamiltonian formulations can be brought together in the framework of the
covariant theory of symplectic structure. They can be identified as
different components of the covariant Witten-Zucker\-man symplectic
$2$-form current density where a variational formulation of the WDVV
equation that leads to the Hamiltonian operator through the Dirac
bracket is available.

\section{Equations of associativity}

In $2$-d topological field theory Witten \cite{witten}, \cite{witten2} has
shown that model independent $n$-point correlation functions follow
recursively from the $2$- and $3$-point correlation functions that serve to
define a non-degenerate flat metric and structure functions of a Frobenius
algebra. These are the principal objects in this theory and they can be
expressed as third derivatives of a generating function 
\begin{equation}
\label{cijk}c_{ijk}=\frac{\partial ^3F}{\partial t^i\,\partial t^j \,
  \partial t^k}
\end{equation}
where $F$ is the free energy \cite{dvv}. One of the independent variables,
say $t^1$, is singled out to serve in the definition of the metric 
\begin{equation}
\label{etaij}c_{1ij}\equiv \eta _{ij}
\end{equation}
which is assumed to be non-degenerate. The structure functions satisfy
a Frobenius algebra and the conditions of associativity
\begin{equation}
\label{assc}c_{mi[j}\,\eta ^{mn}\,c_{k]ln}=0
\end{equation}
result in third order Monge-Amp\`ere equations which are known as WDVV
equations. Indices enclosed by square parantheses are skew-symmetrized.

Dubrovin \cite{dubrovin} has given a systematic account of WDVV equations of
associativity which are completely integrable systems. They admit
bi-Hamilton\-ian structure \cite{fgmn}, \cite{kn} which also provides proof
of their complete integrability through the theorem of Magri \cite{magri}.
However, in the discussion of Hamiltonian structure time, which can be
identifined to be any one of $t^i, i \ne 1$, is necessarily singled out
whereas in the general theory of WDVV equations no such distinction exists.
To give an example, consider the free energy 
\begin{equation}
\label{f1}F_1 = \frac{1}{2} (t^{1})^{2} \, t^{2} + \frac{1}{2} t^{1} \,
(t^{3})^{2} + f( t^{2}, t^{3} )
\end{equation}
that through the identification
\begin{equation}
\label{id1}t^2 = x, \hspace{1 cm} t^3 = t
\end{equation}
results in Dubrovin's equation of associativity
\begin{equation}
\label{dub1}f_{ttt} \, + f_{xxx} \, f_{ttx} - f_{txx}^{\;\;\;\;2} = 0.
\end{equation}
By a trivial interchange in the roles of $t$ and $x$, alternatively if
in place of (\ref{id1}) we make the identification
\begin{equation}
\label{id2}t^2 = t, \hspace{1 cm} t^3 = x, 
\end{equation}
then we are led to 
\begin{equation}
\label{dub2}f_{ttt} \, f_{txx} - f_{ttx}^{\;\;\;\;2} + f_{xxx} = 0 
\end{equation}
which should be the same equation of associativity as there is no
distinction between the independent variables $t$ and $x$ that stemms
from $2$-d topological field theory itself.
However, from the point of view of Hamiltonian
structure Eqs.(\ref{dub1}) and (\ref{dub2}) are radically different. The
bi-Hamiltonian structure of Eq.(\ref{dub1}) which is based on the results of 
\cite{mf} and \cite{gn} was given in \cite{fgmn} and in this paper we shall
present the bi-Hamiltonian structure of Eq.(\ref{dub2}).
We shall find that the Hamiltonian operators appropriate to eq.(\ref{dub2})
are quite different from those obtained for
eq.(\ref{dub1}), however, we shall further show that
they can be identified as different aspects
of the same structure when we consider them in the framework of the covariant
Witten-Zuckerman \cite{cwz} formulation of symplectic structure.
Considerations of covariant symplectic structure take the
variational formulation as their starting point. For WDVV equations
only the Lagrangian that yields their second Hamiltonian structure
through Dirac's theory of constraints \cite{dirac} is known and
therefore this part of our discussion will necessarily be
restricted. We shall show that the second
Hamiltonian operators for eq.(\ref{dub1}) and (\ref{dub2}) are simply the
inverse of the $t$ and $x$ components of the Witten-Zuckerman symplectic
current $2$-form for only one of these equations.

Exactly the same situation holds for the WDVV equation that follows from the
free energy 
\begin{equation}
\label{f2}
F_2 = \frac{1}{2} (t^{1})^{2} \, t^{3} + \frac{1}{2} t^{1} \,
(t^{2})^{2} - \frac{1}{2} (t^{1})^{2} \, t^{2} + f( t^{2}, t^{3} )
\end{equation}
which through the identification (\ref{id1}) leads to
\begin{equation}
\label{univ}
f_{ttt} \, + f_{ttt} \, f_{xxx} - f_{ttx} \, f_{txx} + f_{ttt}
\, f_{txx} - f_{xtt}^{\;\;\;2} + f_{xxx} \, f_{xtt} - f_{xxt}^{\;\;\;2} = 0
\end{equation}
and the bi-Hamiltonian structure of this equation was presented in \cite{kn}.
With the same free energy the opposite identification (\ref{id2}) results in
\begin{equation}
\label{univ2}
f_{ttt} \, f_{xxx} - f_{ttx} \, f_{txx}
+ f_{xxx} \, f_{xtt} - f_{xxt}^{\;\;\;2}
+ f_{ttt} \, f_{txx} - f_{xtt}^{\;\;\;2} + f_{xxx} = 0
\end{equation}
as the WDVV equation of associativity. Once again we shall find that the
bi-Hamiltonian structure of eq.(\ref{univ2}) is different from the
earlier results obtained for eq.(\ref{univ}) but they can be recognized
as different components of the closed conserved current $2$-form
in the covariant formulation of symplectic structure.

\section{System of evolution equations}

  In earlier literature \cite{fors} Monge-Amp\`{e}re equations which are
second order partial differential equations consisting of an appropriate sum
of linear terms and the Hessian were discussed in the
framework of linear equations proper. The reason for this lies in the fact
that the initial value problem for hyperbolic Monge-Amp\`{e}re equations is
identical to that of the second order linear equation \cite{ch}
in spite of the severe non-linearities introduced by the Hessian.
WDVV equations of associativity are of third order but they consist of
a sum of linear terms and the minor determinants of the third order Hankelian
\begin{equation}
\label{hankel}H(u,t,x) = detminor \left\{
\begin{array}{ccc}
f_{ttt} & f_{ttx} & f_{txx} \\ 
f_{ttx} & f_{txx} & f_{xxx} 
\end{array}   \right\}
\end{equation}
closely analogous to the case of Monge-Amp\`{e}re equations. Therefore it
seems reasonable to conjecture that the initial value problem for WDVV
equations is qualitatively the same as the third order linear equation.
This is particularly important in the discussion of the Hamiltonian
structure of WDVV equations where we shall use the techniques originally
developed for real Monge-Amp\'ere equations \cite{ns}. For this purpose
we need to cast the WDVV equation into the form of a triplet of evolution
equations. Introducing the usual auxiliary variables
\begin{equation}
\label{abcdef}a = f_{xxx}, \;\;\;\;\;\;\; b = f_{xxt}, \;\;\;\;\;\;\;  c =
f_{xtt}, 
\end{equation}
we have in place of the WDVV equation (\ref{dub2}) the set of evolution
equations 
\begin{equation}
\label{abceqs}
\begin{array}{rll}
a_t & = & b_x \, , \\ 
b_t & = & c_x \, , \\ 
c_t & = & e_x \, , \\ 
[2mm] e & \equiv & \frac{\textstyle{c^2 - a }}{\textstyle{b}} 
\end{array}
\end{equation}
which consist of equations of hydrodynamic type \cite{dn}. We note that such
a decomposition of eq.(\ref{dub2}) is not unique but this particular choice
of auxiliary variables is useful because the Hamiltonian structure of
equations of hydrodynamic type is a well developed subject. In the
case of eqs.(\ref{abceqs}) the system is linearly degenerate and it will be
necessary to use the results of Ferapontov \cite{f1} on the Hamiltonian
structure of non-diagonizable equations of hydrodynamic type for which
Riemann invariants do not exist. From eqs.(\ref{abceqs}) it is manifest
that $a, b, c$ are conserved densities and there are two others
\begin{equation}
\label{abchams}
\begin{array}{lll}
{\cal P} & = & b \, D^{-1} c, \\
{\cal E} & = & c\, D^{-1} b \, D^{-1} c + D^{-1} a \, D^{-1} b
\end{array}
\end{equation}
which consist of momentum and energy. There exist no further conserved
quantities of hydrodynamic type as eqs.(\ref{abceqs}) are non-diagonizable.
Here and in the following $D^{-1}$ stands for the inverse of the total
derivative operator $D = d/dx$ and its precise definition can be found
in \cite{fokas}. 

\section{Bi-Hamiltonian structure}

    We shall first state the principal result we shall present about
eq.(\ref{dub2}).
\vspace{1mm}

\noindent
{\bf Theorem 1}:

   Eqs.(\ref{abceqs}) can be written as a bi-Hamiltonian system
\begin{equation}
\label{biham}
J_{0} \, \delta H_0 =J_{1} \, \delta H_1
\end{equation}
where $\delta$ denotes the variational derivative with respect to $a, b, c$
and the Hamiltonian operators given by
\begin{equation}
\label{j0abc}  J_0 =   \left(
\begin{array}{ccc}
2 a D + 2 D a & 3 b D + b_{x} & 2 c D \\ 
3 b D + 2 b_{x} & c D + D c & e D \\ 
2 D c & D e & - 3 D
\end{array}           \right)  ,
\end{equation}
\begin{equation}
\label{j1}
J_{1} =  \left(  \begin{array}{ccc}
- D^{3} & 0 & 0 \\ 
0 & 0 & D^{2} 
\frac{\textstyle{1}}{\textstyle{b}} D \\ [2mm] 0 & D \frac{\textstyle{1}}{%
\textstyle{b}} D^{2} & D \frac{\textstyle{1}}{\textstyle{b}} D \frac{%
\textstyle{c}}{\textstyle{b}} D + D \frac{\textstyle{c}}{\textstyle{b}} D 
\frac{\textstyle{1}}{\textstyle{b}} D 
\end{array} \right)
\end{equation}
are compatible so that according to Magri's theorem \cite{magri}
we have a completely integrable system.

   The Hamiltonian functions that yield the equations of motion are
integrals of the densities ${\cal H}_0 = b $ and ${\cal H}_1 = {\cal E}$
respectively.

\section{Spectral problem}

    The general framework for casting WDVV equations into the form
of a spectral problem was given by Dubrovin \cite{dubrovin}. In the
case of eq.(\ref{dub2}), or rather eqs.(\ref{abceqs}) these considerations
yield the Lax pair
\begin{equation}
\label{spectral}
\begin{array}{c}
\Psi _x = z A \Psi,  \hspace{1cm} \Psi _t = z B \Psi, \\[3mm]
A = \left( \begin{array}{ccc}
0 & 1 & 0 \\ a & 0 & b \\ b & 0 & c \end{array} \right),  \hspace{1cm}
B = \left( \begin{array}{ccc}
0 & 0 & 1 \\ b & 0 & c \\ c & 1 & e \end{array} \right)
\end{array}
\end{equation}
where $z$ is the spectral parameter and the compatibility conditions
$$ A_t = B_x, \hspace{1cm} [A,B]=0 $$
are satisfied by virtue of eqs.(\ref{abceqs}).
From the compatibility conditions it follows that
the roots $u^1, u^2, u^3$ of the characteristic equation
\begin{equation}
\label{charac}\det ( \, \lambda I - A \, ) = \lambda^3  - c \lambda^2
- a \lambda + ac - b^2   = 0
\end{equation}
are conserved Hamiltonian densities for eqs.(\ref{abceqs}).

\section{First Hamiltonian structure}

    The easiest way for finding the first Hamiltonian structure of
eqs.(\ref{abceqs}) is to transform to a set of new dependent variables
which consist of the roots of the cubic (\ref{charac}) because the
Hamiltonian operator then assumes the form of a first order homogeneous
operator with constant coefficients. The relationship between the roots of
this cubic to the hydrodynamic variables $a, b, c$ is given by
\begin{equation}
\label{viete}
\begin{array}{rcl}
a & = & - \beta, \\
b & = & \mp  \sqrt{\gamma - \alpha \beta },  \\
c & = & \alpha,
\end{array}
\hspace{1cm}
\begin{array}{rcl}
\alpha & = & u^1 + u^2 + u^3, \\
\beta & = & u^1 u^2 + u^2 u^3 + u^3 u^1, \\
\gamma & = & u^1 u^2 u^3
\end{array}
\end{equation}
according to the formulas of Vi\`ete.
In these variables the equations of motion (\ref{abceqs}) assume the form
\begin{equation}
\label{u123}u^{i}_{t} =  \left(
\frac{\textstyle{(u^{i})^2+\beta }}{\textstyle{\sqrt{\gamma - \alpha \beta}}}
 \right)_x  , \hspace{1cm}  i=1,2,3
\end{equation}
of Hamilton's equations with the Hamiltonian operator
\begin{equation}
\label{j0t}
J_0 ={\frac{1}{2}} {\pmatrix {1&-1&-1\cr -1&1&-1 \cr -1&-1&1 \cr}} D
\end{equation}
and the Hamiltonian density 
$$ {\cal H}_0 = b = \sqrt{u^1 u^2 u^3 - ( u^1 + u^2 + u^3 )
   ( u^1 u^2 + u^2 u^3 + u^3 u^1 ) } .  $$
A lengthy but straight-forward transformation of this operator into
the original auxiliary variables yields the result in eq.(\ref{j0abc}).
The verification of the Jacobi identities for Hamiltonian operator
(\ref{j0abc}) is straight-forward. For this operator the conserved
quantity $\frac{1}{2}a$ results in the trivial flow while $c$ is a Casimir.

\section{Variational principle}

    The most direct way of obtaining the second Hamiltonian structure
of eq.(\ref{dub2}) is through the construction
of a variational principle. The auxiliary variables $a, b, c$ are not
best suited for this purpose, instead it turns out that in terms of
\begin{equation}
\label{pqrdef}
p = f_{x}\, , \;\;\;\;\;\;\; q = f_{t}\, , \;\;\;\;\;\;\; r =
f_{tt}\, , 
\end{equation}
the Lagrangian is given by a simple local expression. The equations of
motion are
\begin{equation}
\label{pqreqs}
\begin{array}{lll}
p_t & = & q_x \, , \\ 
q_t & = & r \, , \\ 
[2mm] r_t & = & \frac{\textstyle{r_{x}^{\,\;2} - p_{xx}}}{\textstyle{q_{xx}}}
\end{array}
\end{equation}
and it can be readily verified that the variational principle with the
Lagrangian density
\begin{equation}
\label{lag}
{\cal L} = \frac{1}{2} p_{x} \, p_{t}  + q_{x} \, r_{x} \, q_{t}
- \frac{1}{2} q_{x}^2 \, r_{t} + p \, q_{xx} + \frac{1}{2} r^2 q_{xx}
\end{equation}
yields eqs.(\ref{pqreqs}). We note that this Lagrangian is linear in
the velocities so that the Hessian vanishes identically and we have a
degenerate Lagrangian system. The passage to its Hamiltonian
formulation requires the use of Dirac's theory of constraints \cite{dirac}.

\section{Dirac bracket}

   The Dirac bracket which replaces the Poisson bracket for systems subject
to constraints plays a central role in the construction of the Hamiltonian
operator for integrable systems \cite{ynz}. This construction is directly
applicable to the WDVV equation (\ref{dub2}) as we shall now detail. Since
the Lagrangian (\ref{lag}) is degenerate the canonical momenta cannot
be inverted for the velocities and following Dirac we introduce the
definition of the momenta
\begin{equation}
\label{constraints}
\begin{array}{ll}
\phi_1 = & \pi_{p} - 
\frac{1}{2} p_{x} \, , \\ \phi_2 = & \pi_{q} - q_{x} \, r_{x} \, , \\ 
\phi_3 = & \pi_{r} + \frac{1}{2} q_{x}^{\;2} 
\end{array}
\end{equation}
as primary constraints. Calculating the Poisson bracket of the constraints
\begin{equation}
\label{pbconst}
\begin{array}{lll}
\{ \phi_{1} (x) , \phi_{1} (y) \} & = & \frac{1}{2} \delta_{x} (y-x)
- \frac{1}{2} \delta_{y} (x-y) \, , \\
 \{ \phi_{1} (x) , \phi_{2} (y) \} & = & 0 \,, \\
\{ \phi_{1} (x) , \phi_{3} (y) \} & = & 0 \, , \\
\{ \phi_{2} (x) , \phi_{2} (y) \} & = & r_{x} \, \delta_{x} (y-x) - r_{y} \,
\delta_{y} (x-y) \, , \\ 
\{ \phi_{2} (x) , \phi_{3} (y) \} & = & q_{x} \, \delta_{x} (y-x) + q_{y} \,
\delta_{y} (x-y) \, , \\ 
\{ \phi_{3} (x) , \phi_{3} (y) \} & = & 0   \,,
\end{array}
\end{equation}
we find that the constraints (\ref{constraints}) are second class as
they do not vanish modulo the constraints themselves. This is the
case with almost all completely integrable systems.
The total Hamiltonian density of Dirac is given by
\begin{equation}
\label{h0}
{\cal H}_{T} = {\cal H}_1  + \Sigma_{i=1}^3 c^i \, \phi_i,  \;\;\;\;\;
{\cal H}_1 = - \frac{1}{2} r^2 q_{xx} - p \, q_{xx}
\end{equation}
where $c^i$ are Lagrange
multipliers. The expression for ${\cal H}_1$ is obtained from the
Lagrangian (\ref{lag}) by Legendre transformation.
The conditions that the constraints are maintained in time
$ \{ \phi_i (x) , H_{T} \} = 0$
give rise to no further constraints but rather determine
the Lagrange multipliers
$$ c^1 = q_{x} \, , \hspace{1cm} c^2 = r \, , \hspace{1cm}
c^3 = \frac{r_{x}^{\;2} - p_{xx}}{q_{xx}}  $$
and we have no secondary constraints.
From eq.(\ref{h0}) we find that the total Hamiltonian is given by
\begin{equation}
{\cal H}_{T}  =   q_{x} \, \pi_{p} + r \, \pi_{q} +
\frac{\textstyle{\ r_{x}^{\;2} - p_{xx}}}{\textstyle{q_{xx}}}
\left( \pi_{r} + \frac{1}{2} q_{x}^{\;2} \right) + \frac{1}{2} p_{x} \, q_{x}
\end{equation}
in terms of the full set of canonical variables. For systems subject to
second class constraints we can solve the constraints to eliminate the
canonical momenta because in Dirac's theory second class constraints hold
as strong equations. As a result the total Hamiltonian density is simply
${\cal H}_1$ which up to a total derivative is the same as ${\cal E}$ in
eq.(\ref{abchams}).

   Given any two smooth functionals $A, B$ the Dirac bracket is defined by
\begin{equation}
\label{diracbr}   \begin{array}{rcl}
\{ A(x) , B(y) \}_{D} & = &   \{ A(x) , B(y) \}  \\[2mm]   & &  -
\int  \{ A(x) , \phi_{i}(z) \} \, J^{ik}(z,w) \, \{ \phi_{k}(w) , B(y) \}
\, d z \, d w            \end{array}
\end{equation}
where $J^{ik}$ is the inverse of the matrix of Poisson brackets of the
constraints. From the definition of the inverse
$$ \int \{ \phi_{i} (x) , \phi_{k} (z) \} J^{kj} (z,y) d z =
\delta_{i}^{j} \, \delta ( x - y ) $$
we end up with a set of differential equations for $J^{ik}$ which can be
solved to yield
\begin{equation}
\label{mm}
\begin{array}{ll}
J^{11} (x, y) = & - \theta (x-y) \, , \\
[2mm] J^{12} (x, y) = & 0 \, , \\ 
[2mm] J^{13} (x, y) = & 0 \, , \\ 
[2mm] J^{22} (x, y) = & 0 \, , \\ 
[2mm] J^{23} (x, y) = & - 
\frac{\textstyle{1}}{\textstyle{q_{xx}}} \, \delta (x-y) \, , \\ [3mm]
J^{33} (x, y) = & - 2 \frac{\textstyle{r_{x}}} {\textstyle{q_{xx}^{\;\;2}}}
\delta_{x} (x-y) + \frac{\textstyle{r_{x} \, q_{xxx}}} {\textstyle{%
q_{xx}^{\;\;3}}} \delta (x-y)
\end{array}
\end{equation}
where the Heaviside unit step function is denoted by $\theta$.

\section{Second Hamiltonian structure}

The transition from the Dirac bracket to the Hamiltonian operator is given
by 
\begin{equation}
\label{defhamop}
\{ u^{i}(x) , u^{k}(y) \}_{D} = - J^{ik}(x,y) \equiv - J^{ik}(x) \delta(x-y)
\end{equation}
since the Poisson brackets of $u^i$ vanish and the constraints are
linear in the momenta.
From eqs.(\ref{defhamop}) and (\ref{mm}) it follows that the Hamiltonian
operator corresponding to the degenerate Lagrangian (\ref{lag}) is simply
\begin{equation}
\label{j1pp}
J_1= - \left(
\begin{array}{ccc}
 D^{-1} & 0 & 0 \\
0 & 0 & \frac{\textstyle{1}}{\textstyle{q_{xx}}} \\ [2mm]
0 & - \frac{\textstyle{1}}{\textstyle{q_{xx}}} &
 \frac{\textstyle{1}}{\textstyle{q_{xx}}} D
\frac{\textstyle{r_x}}{\textstyle{q_{xx}}}
+ \frac{\textstyle{r_x}}{\textstyle{q_{xx}}} D
\frac{\textstyle{1}}{\textstyle{q_{xx}}}
\end{array}
\right) 
\end{equation}
and the proof of the Jacobi identities for the Hamiltonian operator
(\ref{j1pp}) follows from the fact that according to eq.(\ref{defhamop})
it is simply a reformulation of the Dirac bracket for which there is a
general proof of the Jacobi identities \cite{pamd}.
The Hamiltonian operator (\ref{j1pp}) looks non-local but this
is only superficial and related to the choice of variables $p, q$ and $r$.
If we revert to the original auxiliary variables (\ref{abcdef})
$$a=p_{xx} \, , \, \, \, \, b=q_{xx} \, , \, \, \, \, c=r_{x} \, ,$$
the Hamiltonian operator (\ref{j1pp}) is transformed to the form (\ref{j1})
which is a local homogeneous third order operator of hydrodynamic type
that was studied in \cite{doyle}, \cite{potemin}.

\section{Symplectic representation}

    The symplectic formulation of eqs.(\ref{pqreqs}) that provides the
dual description to the Hamiltonian operator (\ref{j1pp}) requires
the inverse of this operator. However, we have just seen that this
Hamiltonian operator is derived from the Dirac
bracket which in turn was obtained from the inverse of the Poisson bracket
of the constraints. The matrix of symplectic $2$-form density
can therefore be obtained directly from eqs.(\ref{pbconst}) and we get
\begin{equation}
\label{oij}
\omega_{ij} = - \left(
\begin{array}{ccc}
 D & 0 & 0 \\
0 & D r_x + r_{x} D & - q_{xx} \\
0 &  q_{xx}  & 0
\end{array}
\right) 
\end{equation}
which can be verified to be the inverse of (\ref{j1pp}). The
symplectic $2$-form density is then given by
\begin{equation}
\label{omu}
\omega = - \frac{1}{2} d p \wedge d p_{x}  - r_{x} \, d q \wedge
d q_{x} + q_{xx} \, d q \wedge d r
\end{equation}
and since this $2$-form is closed, using Poincar\'e's lemma we can write
\begin{equation}
\label{amu}
\begin{array}{ll}
\omega = & d \alpha, \\[1mm]
\alpha = & \frac{1}{2} p_{x} \, d p - r \, q_{xx} \, d q
\end{array}
\end{equation}
in a local neighborhood. In terms of $f$ that enters into the original
formulation of the WDVV equation (\ref{dub2}) the
symplectic $2$-form (\ref{omu}) reduces to
\begin{equation}
\label{omuf}
\omega =  2 \, f_{tx} \, d f_{tt} \wedge d f_{tx}
                  -  \frac{1}{2} \, d f_{x} \wedge d f_{xx}
\end{equation}
using the definitions (\ref{pqrdef}) and discarding a total derivative.

\section{Darboux's theorem}

    The transformation of the third-order Hamiltonian operator (\ref{j1}) to
the canonical form
\begin{equation}
\label{constj1}
  J_{1} = - \left( \begin{array}{ccc}
0 & 0 & 1 \\ 0 & 1 & 0 \\ 1 & 0 & 0 \end{array}
\right) D 
\end{equation}
required by an as yet unproved generalization of Darboux's theorem for third
order Hamiltonian operators is usually given by a differential substitution
where the Casimirs of $J_1$ play a central role.
For the Hamiltonian operator (\ref{j1}) the Casimirs are
\begin{equation}
\label{casimirs}
\begin{array}{lll}
s^1 & = & D^{-1} b , \\ 
[1mm] s^2 & = & D^{-1} a , \\ 
[1mm] s^3 & = & b \, D^{-1} c 
\end{array}
\end{equation}
and the differential substitution required by Darboux's theorem is
obtained by inverting eqs.(\ref{casimirs}). Indeed it can be readily
verified that under this transformation of variables the second Hamiltonian
operator (\ref{j1}) is transformed to the form (\ref{constj1})
which is proof of Darboux's theorem for this case. Finally we note that in
terms of the Casimir variables the equations of motion assume the form
\begin{equation}
\label{heat}
\begin{array}{lll}
s^{1}_{\;t} & = & \left( 
\frac{\textstyle{s^3 }}{\textstyle{s^{1}_{\;x} }}  \right)_x \\ [4mm]
s^{2}_{\;t} & = & s^{1}_{\;x} \\ 
s^{3}_{\;t} & = & \left[ \frac{\textstyle{s^3}}{\textstyle{s^1_{\;x} }}
\left(\frac{\textstyle{s^3 }}{\textstyle{s^1_{\;x} }} \right)_x - s^2
\right]_x 
\end{array}
\end{equation}
of an integrable coupled dispersive system.

\section{Summary of earlier results}

    In order to compare the results on bi-Hamiltonian formulations
of eqs.(\ref{dub2}) and (\ref{dub1}) we need to summarize the
results of \cite{fgmn} on eq.(\ref{dub1}). In this case
\begin{equation}
\label{oldabceqs}
e \equiv  a c - b^2
\end{equation}
replaces the definition the same quantity in eqs.(\ref{abceqs}).
The Hamiltonian operators are given by
\begin{equation}
\label{j0dub1}
{J'}_0 = \left(      \begin{array}{ccc}
-\frac{3}{2} D & \frac{1}{2} D a & D b \\
[2mm] \frac{1}{2} a D & \frac{1}{2} ( D b + b D ) & \frac{3}{2} c D + c_x \\
[2mm] b D & \frac{3}{2} D c - c_x & (b^2 -ac) D + D (b^2 -ac)
\end{array}  \right) \, ,
\end{equation}
\begin{equation}
\label{j1dub1}
{J'}_{1} = \left(  \begin{array}{ccc}
0 & 0 & D^{3} \\ 
[2mm] 0 & D^{3} & - D^{2} a \, D \\ 
[2mm] D^{3} & - D \, a \, D^{2} & {
\begin{array}{c}
D^{2} b \, D + D \, b \, D^{2} \\ 
+ D \, a \, D \, a \, D 
\end{array}   }  \end{array}    \right) \, ,
\end{equation}
and the corresponding densities of Hamiltonian functions
\begin{equation}
{{\cal H}'}_0 = c \, , \hspace{1cm} {{\cal H}'}_1 = - \frac{1}{2} \, a \,
(D^{-1}b)^2 - (D^{-1}b) \, (D^{-1}c)
\end{equation}
yield the system (\ref{abceqs}) with this redefinition of $e$.
These Hamiltonian operators are also compatible.
The symplectic $2$-form obtained from the inverse of (\ref{j1dub1})
is given by
\begin{equation}
\omega' = d p \wedge d r - q_{xx} \,  d p \wedge d p_x  + p_{xx} \,
 d p \wedge d q_{x}    + \frac{1}{2} d q \wedge d q_{x}
\label{omudub1}
\end{equation}
which in terms of $f$ in eq.(\ref{dub1}) is simply
\begin{equation}
\label{omufdub1}
\omega' = 2 \, f_{tt} \, d f_{tx} \wedge d f_{tt}
                  +  \frac{3}{2} \, d f_{x} \wedge d f_{tx}
\end{equation}
where we have again discarded a total derivative.

\section{Covariant formulation}

    We have noted at the beginning that eqs.(\ref{dub2}) and (\ref{dub1})
are obtained by a simple flip of the independent variables. However,
any comparison of the Hamiltonian operators (\ref{j0dub1}) and (\ref{j1dub1})
with (\ref{j0abc}) and (\ref{j1}) yields nothing more than a complete
mismatch even though they arise from what is in fact the same WDVV equation.
The fact that the Hamiltonian operators for these equations look very
different forces us to look for a unifying framework which exists in the
covariant formulation of symplectic structure \cite{cwz}. Indeed these
different looking Hamiltonian structures are simply different components
of the Witten-Zuckerman closed, conserved current $2$-form. In order to show
that this is indeed the case we start with the Lagrangian for, say
eq.(\ref{dub2}),
\begin{equation}
\label{lag2}
{\cal L} = \frac{1}{2} f_{tt}^{\;\;2} \, f_{txx}  - f_{tx}^{\;\;2} \, f_{ttt}
- \frac{1}{2} f_{xx}\, f_{tx}
\end{equation}
expressed in terms of the original variable $f$.
This is the same as the Lagrangian (\ref{lag}) up to a total derivative.
The Witten-Zuckerman current $2$-form $\omega^{\mu}$ which is closed
and conserved
\begin{equation}
\label{wz}
 \omega^{x}_{\;x} + \omega^{t}_{\;t} = 0, \hspace{2cm}
  d \, \omega^{\mu} = 0,  \hspace{5mm} \mu = x, t
\end{equation}
can be obtained directly from the Lagrangian (\ref{lag2}). We find
\begin{equation}
\label{omdub2} \begin{array}{rcl}
 \omega^{x} & = & 2 \, f_{tt} \, d f_{tx} \wedge d f_{tt}
                  +  \frac{3}{2} \, d f_{x} \wedge d f_{tx} , \\[2mm]
 \omega^{t} & = & 2 \, f_{tx} \, d f_{tt} \wedge d f_{tx}
                  -  \frac{1}{2} \, d f_{x} \wedge d f_{xx}
\end{array}
\end{equation}
which satisfies the properties of the symplectic $2$-form listed in
eqs.(\ref{wz}). If we go back to eqs.(\ref{omuf}), (\ref{omufdub1}) for the
symplectic $2$-forms appropriate to eqs.(\ref{dub2}), (\ref{dub1})
expressed in terms of $f$ alone, then we find simply
\begin{equation}
\label{omid}
 \omega^{x} = \omega', \hspace{1cm} \omega^{t} = \omega ,
\end{equation}
as we should expect, since these WDVV equations are related by a flip of
$t$ and $x$. Unfortunately the lack of a variational formulation of WDVV
equations (\ref{dub2}), or (\ref{dub1}) that yields their first Hamiltonian
structure through Dirac's theory of constraints makes it impossible to
present a similar covariant description of their first symplectic structure.

\section{Second pair of WDVV equations}

    In order to discuss the Hamiltonian structure of the WDVV equation
(\ref{univ2}) we need to redefine
\begin{equation}
\label{2abceqs}
e \equiv \frac{b^2 + c^2 + b c - a c - a}{a+b}
\end{equation}
in the system of evolution equations (\ref{abceqs}). The bi-Hamiltonian
structure of these equations of hydrodynamic type is obtained through
the same process we presented earlier and the result can be summarized
in the following.
\vspace{1mm}

\noindent
{\bf Theorem 2}:

   The equations of hydrodynamic type (\ref{abceqs}) with $e$ given by
(\ref{2abceqs}) admit bi-Hamiltonian structure (\ref{biham}) with
Hamiltonian operators
\begin{equation}
\label{2j0abc}
\tilde{J}_0 = \left(
\begin{array}{ccc}
- a D - D a & \frac{1}{2} D ( a - b ) - b D & D b - c D \\[1mm]
\frac{1}{2} (a - b ) D - D b & \frac{1}{2} (b - c ) D + \frac{1}{2} D ( b
-c) & \frac{1}{2} (c - e) D + D c \\[2mm] b D - D c &
\frac{1}{2} D ( c - e) + c D &  \frac{3}{2} D  + D e + e D
\end{array}   \right) ,
\end{equation}
\begin{equation}
\label{2j1}
\tilde{J}_{1} = \left(
\begin{array}{ccc}
- D^{3} & D^{3} & - D^{3} \\ 
D^{3} & - D^{3} & D^{2} 
\frac{\textstyle{a+b+1}}{\textstyle{a+b}} D \\ [2mm] - D^{3} &
D \frac{\textstyle{a+b+1}}{\textstyle{a+b}} D^{2} &
\begin{array}{c} D \frac{\textstyle{1}}{\textstyle{a+b}}
D \frac{\textstyle{b + c + \frac{1}{2}}}{\textstyle{a+b}} D \\
+ D \frac{\textstyle{b + c + \frac{1}{2} }}{\textstyle{a+b}}
D \frac{\textstyle{1}}{\textstyle{a+b}} D \\
- D \frac{\textstyle{a+b+1}}{\textstyle{a+b}} D 
\frac{\textstyle{a+b+1}}{\textstyle{a+b}} D 
\end{array} \end{array} \right)
\end{equation}
and the Hamiltonian functions which are integrals of the densities
$\tilde{{\cal H}}_0 = c $ and
\begin{equation}
\label{4pqrhams}
\tilde{{\cal H}}_1  =  ( c - b - a )\, D^{-1} b \, D^{-1} c
 + ( b + 1 ) \,D^{-1} a \, D^{-1} b + c D^{-1} a D^{-1} c + \frac{1}{2}
( D^{-1} a )^2
\end{equation}
yield the equations of motion. The Hamiltonian operators
$\tilde{J}_0$ and $\tilde{J}_1$ are compatible so that
the theorem of Magri is applicable in this case as well.

     Comparison of these results with the bi-Hamiltonian structure of
eq.(\ref{univ}) presented in \cite{kn} again results in a mismatch
which, however, is not as severe as before since eqs.(\ref{univ}) and
(\ref{univ2}) resemble each other rather closely.
In order to show that these different Hamiltonian structures simply
correspond to different components of the Witten-Zuckerman $2$-form
first we note that the variational principle with the Lagrangian
\begin{equation}
\label{4lag}
\begin{array}{rl}
\tilde{{\cal L}} = & - \frac{1}{2}  \left( f_{xx}^{\;2} +
   f_{tx} \, f_{xx} \right)
 + f_{xxx} \, f_{tt} \, f_{xx} + f_{txx} \, f_{tx} \, f_{tt} \\[1mm] &
 - f_{xxx} \, f_{tt} \, f_{tx}
 - f_{ttt} \, f_{tx} \, f_{xx} -  \frac{1}{2} \, f_{ttt} \, f_{tx}^{\;\;2}
\end{array}
\end{equation}
yields eq.(\ref{univ2}), or more precisely we get a linear combination of
both $t$ and $x$ derivatives of eq.(\ref{univ2}).
This Lagrangian can be rewritten in terms
of the auxiliary variables (\ref{pqrdef})
\begin{equation}
\label{3lag}
\begin{array}{rl}
\tilde{{\cal L}} = & \left( - q_{x} \, q_{xx}
  + q_{x} \, r_{x} - r \, p_{xx} + p_{x}
\, q_{xx}  + \frac{1}{2} p_{x} \right) \, p_{t}
- \left( \frac{1}{2} q_{x}^2 + p_{x} \, q_{x} \right) r_{t} \\[1mm] &
 + \left(  q_{x} \, r_{x} - q_{x} \, p_{xx}
 + p_{x} \, p_{xx} + p_{x} \, r_{x} \right) \, q_{t}
 - q_{x} \, r \, r_{x} - \frac{1}{2} p_{x}^{\;2}
 \\ [1mm] & + q_{x} \, q_{xx} \, r
+ q_{x} \, p_{xx} \, r - p_{x} \, q_{x} \, q_{xx} - p_{x} \, r \, r_{x}
- q_{x} \, p_{x} 
\end{array}
\end{equation}
which yields a triplet of evolution equations, the integrability conditions
of which result in eq.(\ref{univ2}).

   The symplectic $2$-form obtained from the inverse of the Hamiltonian
operator (\ref{2j1}) is given by
\begin{equation}
\label{omuyeni}
\begin{array}{ll}
\tilde{\omega} = & \left[ \, ( r_x + q_{xx} ) \, d p_{x}
+ ( r_{x} - p_{xx} ) \, d q_{x}
- ( p_{xx} + q_{xx} ) \, d r \, \right] \wedge ( d p + d q ) \\
[1mm] & - \frac{1}{2} d p \wedge d p_{x}
\end{array}
\end{equation}
and this is rather similar to the result 
\begin{equation}
\label{omueski}
\begin{array}{ll}
\tilde{\omega}' = & \left[ \, ( r_x + q_{xx} ) \, d p_{x}
+ ( r_{x} - p_{xx} ) \, d q_{x}
- ( 1 + p_{xx} + q_{xx} ) \, d r \, \right] \wedge ( d p + d q ) \\ 
[1mm]  & + \frac{1}{2} d q \wedge d q_{x}
\end{array}
\end{equation}
for eq.(\ref{univ}) that was obtained earlier in \cite{kn}. The
Witten-Zuckerman symplectic $2$-form that follows from the
Lagrangian (\ref{4lag}) is
\begin{equation}
\label{omjan2} \begin{array}{rcl}
 \omega^{x} & = & 2 \left( f_{txx} - f_{ttx} -\frac{3}{4} \right)
   d f_{tx} \wedge d f_{x}
  - \left( f_{ttx} + 1  \right)  d f_{xx} \wedge d f_{x} \\[2mm] &  &
   + \left( 2\, f_{txx} - f_{xxx} \right)  d f_{tt} \wedge d f_{x}
   + \left( 2\, f_{tt} + d f_{xx} \right) d f_{xx} \wedge d f_{tt} \\[2mm]&&
   + 2 \, f_{tt} \,  d f_{tx} \wedge d f_{tt}
   - f_{ttx} \,  d f_{tx} \wedge d f_{t}
   + f_{txx} \,  d f_{tt} \wedge d f_{t}
   - 2\, f_{tt} \,  d f_{tx} \wedge d f_{xx} \\[2mm]
 & = & \tilde{\omega}'              \\[2mm]
 \omega^{t} & = & \frac{1}{2} d f_{xx} \wedge d f_{x}
      +  \left[ \left( f_{ttx} + f_{txx} \right) d f_{xx}
  + \left( f_{ttx} - f_{xxx} \right)  d f_{tx} \right. \\[2mm] & & \left.
   - \left( f_{xxx} + f_{txx} \right)  d f_{tt} \right]
   \wedge \left( d f_{x} + d f_{t} \right)   \\[2mm]
 & = & \tilde{\omega}
\end{array}
\end{equation}
where $\tilde{\omega}, \tilde{\omega}'$ are given by eqs.(\ref{omueski})
and (\ref{omuyeni}) respectively. This is exactly the same final result
(\ref{omid}) we found earlier for eqs.(\ref{dub2}) and (\ref{dub1}).

\section{Conclusion}

   The bi-Hamiltonian structure of WDVV equations of associativity
is expressed by different pairs of Hamiltonian operators depending
on which independent variable in the free energy is chosen to play
the role of time. However, we have shown
that the symplectic $2$-forms which can be obtained from the second
Hamiltonian operators are simply different components of the covariant
Witten-Zuckerman conserved current symplectic $2$-form. We have been able
to obtain this result only for the second Hamiltonian structure
because only in this
case does there exist a degenerate Lagrangian subject to second class
constraints which through the Dirac bracket yields the Hamiltonian
operator. The lack of a variational principle for the WDVV equations
that yields their different first Hamiltonian structures makes it impossible
to discuss them together in the framework of the covariant theory
of symplectic structure.

\end{document}